\documentclass[reprint,onecolumn,notitlepage,longbibliography]{revtex4-1}

\usepackage{amsmath,amssymb}    
\usepackage{graphicx}
\usepackage{dcolumn}
\usepackage{bm}
\usepackage{caption} 
\usepackage{natbib}
\usepackage{booktabs} 
\graphicspath{{./Fig/}} 

\begin{document}

\title{Effect of cytosol viscosity on the flow behavior of red blood cell suspensions in microvessels}

\author{Wei Chien}
\author{Gerhard Gompper}
\author{Dmitry A. Fedosov}\email{d.fedosov@fz-juelich.de}
\affiliation{Theoretical Soft Matter and Biophysics, Institute of Complex Systems and Institute for Advanced Simulation, Forschungszentrum J\"ulich, 52425 J\"ulich, Germany}

\date{\today}

\begin{abstract}
The flow behavior of blood in microvessels is directly associated with tissue perfusion and oxygen delivery. Current efforts on modeling blood flow have primarily focused 
on the flow properties of blood with red blood cells (RBCs) having a viscosity ratio $C$ of unity between the cytosol and suspending medium, while under physiological 
conditions the cytosol viscosity is about five times larger than the plasma viscosity (i.e., $C\approx 5$). The importance of $C$ for the behavior of single RBCs in fluid flow 
has already been demonstrated, while the effect of $C$ on blood flow has only been sparsely studied. We employ mesoscopic hydrodynamic simulations to perform a systematic 
investigation of flow properties of RBC suspensions with different cytosol viscosities for various flow conditions in cylindrical microchannels. Our main aim is to link 
macroscopic flow properties such as flow resistance to single cell deformation and dynamics as a function of $C$. Starting from a dispersed cell configuration, we find 
that the flow convergence and the development of a RBC-free layer (RBC-FL) depend only weakly on $C$, and require a convergence length in the range of $25D-50D$, where $D$ is the channel diameter. 
The flow resistance for $C=5$ is nearly the same as that for $C=1$, which is facilitated by a slightly larger RBC-FL thickness for $C=5$. This effect is due to the suppression of 
membrane motion and dynamic shape deformations by a more viscous cytosol for $C=5$, resulting in a more compact cellular core of the flow in comparison to $C=1$. 
The weak effect of cytosol viscosity on the flow resistance and RBC-FL explains why cells can have a high concentration of hemoglobin for efficient oxygen delivery, without 
a pronounced increase in the flow resistance. 
\end{abstract}

\maketitle

\section{Introduction}
\label{sec:intro}

Blood is a multi-component suspension which consists of plasma ($\approx 55\%$) and cells (red blood cells $\approx 45\%$, white blood cells and platelets $< 1\%$). 
Flow properties of blood are mainly governed by red blood cells (RBCs) \cite{Popel_MH_2005,Freund_NSB_2014,Fedosov_MBF_2014,Secomb_BFM_2017}, which play an important 
role in many physiological processes. For instance, RBCs are responsible for oxygen delivery and mediate margination (or migration) of platelets 
\cite{zhao11a,Vahidkhah_PDB_2014,Mehrabadi_ESR_2016} and leukocytes \cite{Freund_LMM_2007,Fedosov_WBC_2012,Fedosov_WBC_2014}  toward vessel walls, thus 
affecting the hemostatic process and immune response. Blood-flow properties are also crucial in many applications, such as the enrichment or separation of rare 
circulating tumor cells from blood \cite{Fachin_CTC_2017,Lin_CTC_2018} and the effectiveness of drug carriers for delivery to the targeted sites \cite{Lee_AIP_2013,Mueller_MPB_2014,Cooley_IPS_2018}.
 
Suspensions of blood cells reveal complex flow properties \cite{Secomb_BFM_2017,Pries_BFN_1990} and rheology \cite{Skalak_MBF_1981,Lanotte_RCM_2016}. Two 
representative examples of the flow behavior of RBC suspension in microvessels or glass capillaries are the Fahraeus \cite{Fahraeus_SSB_1929} and Fahraeus-Lindqvist \cite{Fahraeus_VOB_1931,Pries_BVT_1992}  
effects. The former effect concerns RBC volumetric flux (or the so-called discharge hematocrit $H_d$), which appears to be larger than the tube (bulk) hematocrit $H_{t}$ 
in vessels with a diameter $D$ in the range of $7-200$ $\mu m$. The latter effect describes a minimum of blood-flow resistance in a tube with a diameter $\approx 8$ $\mu m$, such that 
the resistance to flow increases for both smaller and larger vessel diameters. The main mechanism governing these phenomena is the formation of a RBC depleted region next to the wall, 
called RBC-free layer (RBC-FL), as the suspension flows \cite{Cokelet_DHD_1991,Goldsmith_RF_1989}. The thickness of RBC-FL is directly associated with blood flow resistance \cite{Reinke_BVS_1987,Sharan_2PM_2001,Fedosov_BFC_2010,Lei_BFT_2013}
and plays a crucial role in the adhesion of leukocytes, platelets, and drug-delivery carriers to vessel walls \cite{Vahidkhah_PDB_2014,Cooley_IPS_2018}. 

The thickness of RBC-FL is governed by a competition between the hydrodynamic lift force acting on RBCs in the direction away from the wall and cell-cell interactions which disperse them and 
drive them toward the wall \cite{Secomb_BFM_2017,Katanov_MVR_2015,Geislinger_HLV_2014}. Particle migration in the Stokes flow regime (i.e., no inertia) relies on breaking 
the time-reversal symmetry, which is achieved through RBC dynamics and/or deformations. For instance, a tumbling rigid spheroidal particle does not experience any migration
on average in the presence of a wall \cite{Pozrikidis_OMS_2005}, but a stable tank-treading motion of a spheroidally shaped membrane leads to the migration 
away from the wall \cite{Olla_LTT_1997,Grandchamp_LSD_2013}. As shown by theoretical analysis \cite{Olla_LTT_1997,Olla_RTT_1997}, 
experiments \cite{Grandchamp_LSD_2013,Abkarian_TTU_2002} and simulations \cite{Messlinger_DRH_2009}, the migration velocity $v_l$ is proportional to 
$f(C)\dot{\gamma}R^{3}/d_w^\alpha$, where $f(C)$ is a function of the ratio $C=\eta_{in}/ \eta_{ex}$ of internal $\eta_{in}$ and external $\eta_{ex}$ fluid viscosities, $\dot{\gamma}$ is the local 
shear rate, $R$ is the characteristic particle size, $d_w$ is the distance away from the wall, and $\alpha$ is an exponent whose value is often reported to be close to two. 
Time reversibility of RBC motion is generally broken by cell orientation and deformation in flow \cite{Geislinger_HLV_2014,Messlinger_DRH_2009,Chen_IDM_2014}.
  
The lift force is counterbalanced by RBC dispersion due to cell-cell hydrodynamic interactions in flow \cite{Secomb_BFM_2017,Katanov_MVR_2015,Kumar_MSF_2012}. 
It is intuitive that fluid flow can significantly enhance such interactions or collisions between particles, which are often referred to as shear-induced dispersion forces 
and depend on local shear rate, particle size, deformation, and dynamics \cite{Grandchamp_LSD_2013,Katanov_MVR_2015,Vollebregt_MPM_2012}.  
The response of RBCs to fluid stresses is known to be sensitive to the viscosity ratio $C$ between internal and external fluids \cite{Lanotte_RCM_2016,Mauer_FIT_2018,Sinha_DRBC_2015,Yazdani_IMV_2013}. 
For example, at low viscosity contrasts $C \lesssim 3$, RBCs tumble at low shear rates and tank-tread at high shear rates \cite{Fischer_SM_2004,Skotheim_RBC_2007,Abkarian_SSF_2007}, 
while for $C \gtrsim 3$, the tank-treading motion is suppressed and replaced by dynamic multi-lobed shapes \cite{Lanotte_RCM_2016,Mauer_FIT_2018}. These differences in RBC deformation 
and dynamics as a function of $C$ are expected to affect the lift force, cell-cell interactions in blood flow, and local structure of RBC suspensions, which influence blood-flow resistance.

The main focus of our investigation is the effect of $C$ on the behavior of RBC suspensions in microvessels and the dependence of RBC-FL thickness and flow resistance
on the viscosity contrast. Even though recognized, the importance of $C > 1$ for blood flow is not well studied so far. Katanov et al.~\cite{Katanov_MVR_2015} have investigated 
the formation of RBC-FL and flow convergence to steady state for $C=1$, starting from an initially dispersed configuration of RBCs, and found that the full flow convergence 
requires a length of about $25D$ ($D$ is the tube diameter), which is nearly independent of flow rate, RBC hematocrit, and channel size for $10$ $\mu m$ $<D<100$ $\mu m$. 
A recent numerical investigation of blood flow in a tube with a diameter of $D=70$ $\mu m$ \cite{deHaan_NIE_2018} has predicted a comparable RBC-FL thickness and flow resistance 
for suspensions with $C=1$ and $C=5$, where the differences become more pronounced at high flow rates. Another simulation study of blood flow in a slit \cite{Saadat_ECV_2019}
has suggested a domination of the lift force on RBCs over cell-cell interactions for $C=1$, such that a slightly smaller RBC-FL was found for $C=5$ suspension in comparison to $C=1$. 
To clarify the importance of $C$ for the behavior of RBC suspensions, we have performed a systematic investigation using mesoscopic hydrodynamic simulations, which include 
suspensions with $C\in[1,20]$ as well as rigidified RBCs, several different flow rates, hematocrits, and tube diameters. In particular, we investigate the development of RBC-FL for the various 
conditions, and connect it to the flow resistance, deformation and dynamics of single RBCs. Our results show that the RBC-FL develops faster for $C=1$ in comparison with $C=5$ 
due to a larger lift force on RBCs with $C=1$. The flow-convergence length becomes larger for elevated $C$ values, but remains within approximately $50D$. The RBC-FL thickness
for $C=5$ is slightly larger than that for $C=1$, resulting in a nearly negligible effect of $C$ on the flow resistance. This property is due to a smaller dispersion of cells for $C=5$ in
comparison to $C=1$, since a larger internal viscosity dampens shape changes and membrane dynamics of RBCs. Suspensions with stiffened RBCs, which approximate the case of 
$C \to \infty$, exhibit the smallest RBC-FL and the largest flow resistance. The robustness of flow resistance with respect to $C\in[1,20]$ permits RBCs to contain a cytosol with 
a high concentration of hemoglobin, which maximizes oxygen delivery and does not strongly affect flow resistance.  
            
The paper is organized as follows. Simulation methods, models, and setup are introduced in Section \ref{sec:method}. Section \ref{sec:results} presents simulation results, where 
the behavior of RBC suspensions with $C\in[1,20]$ is investigated. The analysis of single cell characteristics is performed in Sections \ref{sec:struct} and \ref{sec:shape}, in order 
to explain differences in RBC-FL for various $C$. Finally, the dependence of RBC-FL thickness on several parameters, including flow rate, tube hematocrit and 
diameter, is investigated in Section \ref{sec:other}. Our main results are discussed and summarized in Section \ref{sec:summary}.

\section{Methods and models}
\label{sec:method}

Fluid flow is modeled by the smoothed dissipative particle dynamics (SDPD) method \cite{Espanol_SDPD_2003} with angular momentum conservation \cite{Mueller_SDPD_2015}, which is a mesoscopic particle-based hydrodynamics approach. 
The conservation of angular momentum is crucial for the proper representation of cellular motion when distinct fluid viscosities inside and outside the cell are employed \cite{Mueller_SDPD_2015}. 
RBCs are represented by a spring-network model \cite{Noguchi_STV_2005,Fedosov_SCG_2010,Fedosov_RBC_2010}, and coupled to fluid flow through dissipative forces. Below, we briefly review several model 
ingredients with an emphasis on the implementation of the viscosity contrast between internal and external fluids separated by the membrane. More details about the methods and models can be found in Refs.~\cite{Mueller_SDPD_2015,Fedosov_DDC_2014,Katanov_MVR_2015}.

\subsection{RBC membrane model}
\label{sec:rbc} 

The RBC membrane is represented by a spring-network model \cite{Noguchi_STV_2005,Fedosov_SCG_2010,Fedosov_RBC_2010} with $N_v$ vertices distributed at a biconcave cell shape. Potential energy of the membrane,
\begin{equation}
U_{tot} = U_{sp} + U_{bend} + U_{area} + U_{vol},
\end{equation}
consists of several contributions.  
$U_{sp}$ corresponds to the spring's energy, which mimics elasticity of the spectrin network attached to the back side of the lipid membrane. $U_{bend}$ is the bending energy, representing
bending resistance of the lipid bilayer. $U_{area}$ and $U_{vol}$ impose area and volume conservation constraints, which mimic area incompressibility of the lipid bilayer and incompressibility of a cytosol, respectively. 
The biconcave shape of a RBC at rest is imposed by setting the reduced volume $V^*=6V_r/(\pi D_r^3) = 0.64$, where $V_r$ is the RBC volume and $D_r=\sqrt{A_r/\pi}$, with $A_r$ being the area of a RBC.
The RBC membrane is characterized by the shear modulus $\mu$ and bending rigidity $\kappa$, which are implemented through in-plane elastic forces from the modeled springs and out-of-plane bending forces acting 
on each pair of adjacent triangles, respectively. The membrane parameters are set to mimic average properties of a healthy RBC with $\mu=4.8$ $\mu N/m$ and $\kappa=70k_BT$. The effective size of a RBC is $D_r=6.51$ 
$\mu m$, the surface area is $A_r=132.9$ $\mu m^{2}$, and the total volume is $V_r=92.45$ $\mu m^{3}$. The stress-free shape of a RBC elastic network is assumed to be an oblate spheroid
with a reduced volume of 0.96. 

\subsection{Fluid-membrane interactions}
\label{sec:fm_inter} 

To model the viscosity ratio $C=\eta_{in}/\eta_{ex} \neq 1$, internal and external fluids have to be separated by the membrane. An impenetrable membrane is implemented through bounce-back boundary conditions (BCs) for 
both internal and external fluid particles at every triangular face of the membrane. Thus, internal fluid particles are subject to bounced-back BCs from inside the cell, while external fluid particles are bounced back from the outer 
membrane surface. Different fluid viscosities are implemented through different friction coefficients of dissipative forces in the SDPD method. Dissipative interactions between internal and external fluids assume the average of 
the two friction coefficients. 

The frictional (dissipative) coupling between fluid and membrane particles is implemented through a dissipative force \cite{Fedosov_RBC_2010},   
\begin{equation}
\boldsymbol{f}_{ij}^{D}=\gamma(1-r_{ij}/r_{m})^{\alpha}(\boldsymbol{v}_{ij}\cdot \boldsymbol{e}_{ij})\boldsymbol{e}_{ij}, 	 \quad \quad  r_{ij}=|\boldsymbol{r}_{i}-\boldsymbol{r}_{j}|<r_{m},
\label{eq:coupling}
\end{equation}
similar to that in the dissipated particle dynamics method \cite{Hoogerbrugge_SMH_1992,Espanol_SMO_1995}, 
where $\gamma$ is the friction coefficient, $\alpha=0.2$ is an exponent of the weight function, $r_{m}$ is the cutoff radius, $\boldsymbol{v}_{ij}=\boldsymbol{v}_{i} - \boldsymbol{v}_{j}$ is the velocity difference, and $\boldsymbol{e}_{ij} = \boldsymbol{r}_{ij}/r_{ij}$. 
The value of $\gamma$ is computed as \cite{Fedosov_RBC_2010},  
\begin{equation}
	\gamma = \frac{4A_r \eta}{N_v \rho I_V}, \quad \quad  I_V= \int_{V_h}(1-r_{ij}/r_{m})^{\alpha}zdV,
	\label{eq:gamma}
\end{equation}
where $\eta$ is the fluid viscosity, $\rho$ is the fluid density, and $V_h$ represents a half sphere of radius $r_{m}$ in the positive z direction. This estimation of the friction coefficient assumes 
a linear flow-velocity profile within a distance $r_{m}$ near the membrane surface, so that the local shear rate cancels out. The cutoff for fluid-membrane coupling is set to $r_{m}=0.75$ $\mu m$. 

To prevent an overlap of two membranes, a short-ranged repulsive Weeks-Chandler-Anderson (WCA) potential is applied between pairs of membrane particles belonging to different cells within the cutoff distance $r_{WCA}$. Note that the thickness of 
RBC-FL is slightly sensitive to the choice of $r_{WCA}$, which depends on membrane resolution. For $N_v=1000$, the average spring length at rest is approximately $l_{ave}=0.4$ $\mu m$ and $r_{WCA}$ is set to $0.3$ $\mu m$ 
to prevent overlap. Doubling the membrane resolution allows a slight decrease of $r_{WCA}$, which may result in a slight change of the RBC-FL thickness. Nevertheless, $N_v=1000$ is large enough to properly represent 
membrane deformation and nearly eliminate the effect of the cutoff distance $r_{WCA}$ on blood flow properties.

\subsection{Simulation setup}
\label{sec:setup}

\begin{figure*}[!h]
	\centering
	\includegraphics*[width=0.7\textwidth]{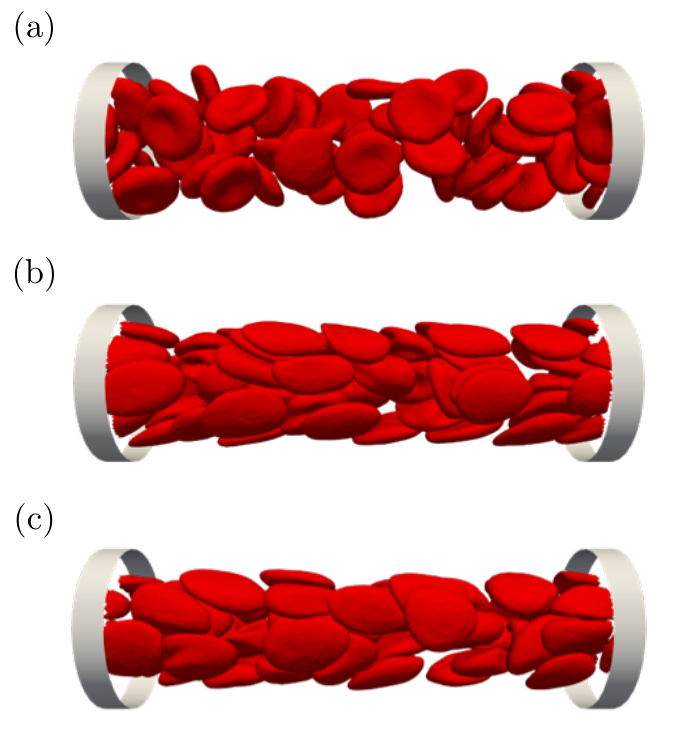}
	\caption{Simulation snapshots for $D=20$ $\mu m$ and $H_{t}=30\%$. (a) Configuration of dispersed RBCs, before the flow is applied. (b) Converged flow of a RBC suspension for $C=1$ and $Ca =  0.61$ ($\overline{\dot{\gamma}} = 377.4$ $s^{-1}$). (c) Converged 
		flow for $C=5$ and the same $Ca$ as in (b). The flow is from left to right.}
	\label{fig:Scheme}
\end{figure*}

The computational domain is a periodic cylindrical tube with a diameter $D$ and a length $L=60$ $\mu m$ ($\approx 9.2D_r$), which is long enough to avoid finite-size effects. Initially, RBCs are introduced into the computational domain with an 
ordered structure. Their number is determined by hematocrit $H_{t}$ (the volume fraction of RBCs in a tube), which is assumed to be $H_t=30\%$ in most simulations. Then, fluid particles are randomly placed with 
a uniform distribution into the computational domain. The number density of fluid particles is $n=9$ $\mu m^{-3}$, the smoothing length for SDPD pair interactions is $r_{c}=1.04$ $\mu m$, resulting in about $30$ 
particles within the interaction range $r_c$. The SDPD fluid particles inside the cells are set to represent the internal fluid with viscosity 
$\eta_{in}$, while particles outside the cells correspond to the external fluid with viscosity $\eta_{ex}$. To relax the initially ordered structure of RBCs, the cellular suspension is mixed by applying a flow within the tube, 
which is driven by a force exerted on all fluid particles. When the cells are well mixed, the flow is stopped, and the RBCs are let to diffuse and fill up the whole tube, resulting in a dispersed RBC configuration illustrated in Fig.~\ref{fig:Scheme}(a).
After the dispersed cell configuration is reached, a constant force $f$ applied on all fluid particles in the x direction is turned on again to drive the fluid flow. Note that this driving force represents a uniform pressure gradient 
$\frac{\Delta P}{L}=n\cdot f$, where $\Delta P$ is the pressure drop over the length $L$. The generated flow leads to RBC migration away from the wall and the formation of a denser cellular region near the tube center, as shown 
in Figs.~\ref{fig:Scheme}(b) and \ref{fig:Scheme}(c) for two different $C$ values.

The flow strength is characterized by the dimensionless capillary number 
\begin{equation}
Ca=\frac{\eta_{ex}\overline{\dot{\gamma}}}{\mu/D_r} = \overline{\dot{\gamma}} \tau,
\end{equation}
which represents the ratio between fluid stresses and elastic membrane stresses. Here, $\overline{\dot{\gamma}}=\overline{v}/D$ is the average shear rate with the average velocity $\overline{v}$ for a Newtonian fluid with viscosity $\eta_{ex}$ driven by the same pressure 
gradient. Thus, the capillary number directly characterizes the applied driving force or the pressure gradient. Note that $\tau=\eta_{ex}D_r/\mu$ represents a characteristic relaxation time of a RBC, which is equal to approximately $1.6\times 10^{-3}$ $s$
for $\mu=4.8$ $\mu N/m$ and blood-plasma viscosity $\eta_{ex}= 1.2$ $mPa\cdot s$. Simulations are performed at low enough Reynolds numbers, such that the largest investigated flow rate corresponds to $Re=\rho \overline{v} D_r/\eta_{ex}=2.34$ ($\rho$ is the fluid density).
Furthermore, the Mach number $Ma=\overline{v}/c_{s}$ with the speed of sound $c_{s}$ is less than $0.1$ in all simulations, so that the fluid flow can be considered incompressible.

Solid wall BCs are modeled by a layer of immobile SDPD particles with a width $r_{c}$. The structure and density of the wall layer are identical to those of equilibrated SDPD fluid in a periodic box. The pair forces between fluid and 
wall particles are the same as those for fluid-fluid interactions. Furthermore, an adaptive shear force is added to fluid particles within a near-wall layer of thickness $r_{c}$ to fully ensure no slip conditions \cite{Fedosov_TDI_2009}. 
To prevent wall penetration, both fluid and membrane particles are reflected back inside the tube domain.

\section{Results}
\label{sec:results}

\subsection{Development of the flow and RBC-FL}
\label{sec:rbc-fl} 

To examine flow development for various viscosity contrasts $C \in [1,20]$, the viscosity of internal fluid $\eta_{in}$ is varied. This range of $C$ covers RBC physiological conditions as well as some diseased states.  After the pressure gradient is applied, 
the flow develops and reaches a terminal velocity profile with a position-averaged velocity $v_{T}$ at long times. Note that the velocity of individual cells depends on their location within the tube and can fluctuate in time even after the average flow velocity has reached steady 
state. $v_{T}$ is inversely proportional to the flow resistance, and can be used to obtain the relative suspension viscosity $\eta_{rel}=\overline{v}/v_{T}$, which compares the volumetric flow rate of RBC suspension with that of external fluid (without RBCs) for the 
same pressure gradient. Thus, $\eta_{rel}$ quantifies an increase in the flow resistance due to the presence of RBCs. The dependence of $\eta_{rel}$ on $C$ for $H_{t}=30\%$ is shown in Table \ref{tab:table1}. Interestingly, the relative viscosity increases
only by about $8\%$ when $C$ is increased $20$ times. Furthermore, Table \ref{tab:table1} also shows $\eta_{rel}=1.7$ for a suspension of stiffened cells (SC), whose shear modulus $\mu$ is $100$ times larger than that of healthy RBCs. Stiffened  
RBCs do not exhibit significant deformation in fluid flow and represent a limit of very large viscosity contrast. 

\begin{table}[]
	\centering
	\caption{Relative suspension viscosity $\eta_{rel}$ and final RBC-FL thickness $\delta_f$ as a function of $C$. $H_{t}=30\%$, $D=20$ $\mu m$, and $Ca =  0.61$ ($\overline{\dot{\gamma}} = 377.4$ $s^{-1}$). 
		"SC" denotes stiffened cells, whose shear modulus is increased $100$ times in 
	comparison to that of a healthy RBC.}
	\begin{tabular}{c|c|c|c|c|c}
		Viscosity contrast ($C$)      & 1    & 5    & 10   & 20   & SC    \\ \hline \hline
		$\eta_{rel}$ & 1.27 & 1.30 & 1.33 & 1.37 & 1.70 \\ \hline
		$\delta_f$ $[\mu m]$ &  2.77 & 3.12 & 3.05 & 3.14 & 2.09 \\ \hline
	\end{tabular}
	\vspace{1ex}		
	\linebreak  
	{\centering  \par}
	\label{tab:table1}%
\end{table}

As the flow develops, RBCs migrate toward the tube center, resulting in the formation of RBC-FL near the wall whose thickness is directly associated with the flow resistance. To measure the thickness of RBC-FL, RBC suspension at a fixed time 
is projected onto the y-z plane \cite{Fedosov_BFC_2010}, which is similar to taking a snapshot from experimental movie \cite{Kim_TSV_2007}. Then, distances between the tube wall and projected RBC-core edge are extracted at several positions 
along the tube. Averaging these distances yields an average thickness of the RBC-FL $\delta$. To improve RBC-FL statistics, we also employ flow axisymmetry, such that the thickness is sampled for $10$ different angles by rotating a simulated configuration 
before the projection onto the y-z plane is performed. Furthermore, the RBC-FL data are accumulated over a certain time window, which is chosen long enough to avoid large deviations in RBC-FL thickness measurements 
and short enough to resolve the dynamics of RBC-FL development. This time window corresponds to about $8$ $ms$, within which RBCs move on average $3$ $\mu m$. 

\begin{figure*}[!h]
	\centering
	\includegraphics*[width=1.0\textwidth]{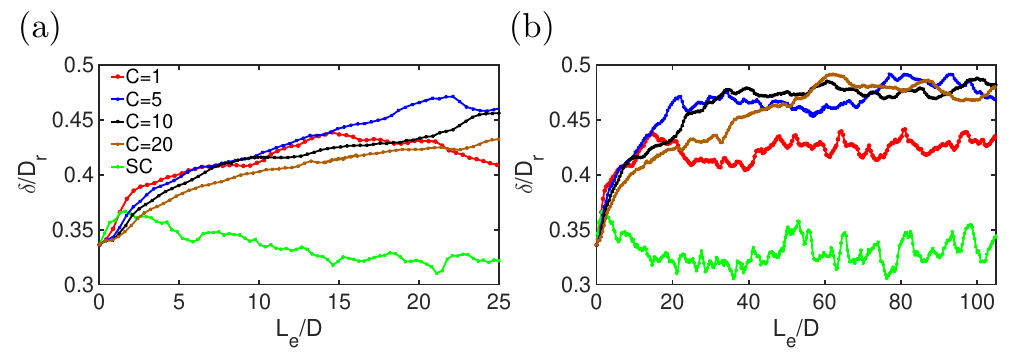}
	\caption{Evolution of the RBC-FL thickness in flow as a function of average convergence length $L_e=\overline{v} t$  for $C=1$, $5$, $10$, $20$ and a suspension of stiffened RBCs. (a) Transient behavior at the beginning of RBC-FL development. 
		(b) Dynamics of the RBC-FL thickness over the total simulation time. $H_{t}=30\%$, $D=20$ $\mu m$, and $Ca =  0.61$ ($\overline{\dot{\gamma}} = 377.4$ $s^{-1}$).}
	\label{fig:CFL}
\end{figure*}

The RBC-FL thickness for the initial configuration without flow (Fig.~\ref{fig:Scheme}(a)) is $\delta \approx 2.2$ $\mu m$. This non-zero RBC-FL thickness is due to finite $H_{t}=30\%$, the biconcave RBC geometry, which affects cell close-packing, and 
entropic repulsion from the wall that originates from the rotational diffusion of RBCs. Figure \ref{fig:CFL} shows the development of RBC-FL as a function of the average flow-convergence length $L_e=\overline{v} t$. At short times ($L_e < 5D = 100$ $\mu m$), $\delta$ 
increases faster for the suspension with $C=1$  than for those with $C > 1$, see Fig.~\ref{fig:CFL}(a). However, the RBC-FL thickness for $C=1$ suspension saturates at a smaller value than $\delta$ for $C> 1$ with a difference of about $200-300$ $nm$, 
as shown in Fig.~\ref{fig:CFL}(b). As $C$ increases from $5$ to $20$, the converged RBC-FL thickness shows a similar plateau value of $\delta \approx 3.1$ $\mu m$. Converged or final RBC-FL thicknesses $\delta_f$ for different $C$ values are given 
in Table \ref{tab:table1}. For the SC suspension with stiffened RBCs, $\delta$ fluctuates within the range of
$2.0-2.3$ $\mu m$, which is close to the RBC-FL thickness without flow. The differences in RBC-FL thicknesses for various $C$ values are due to cell deformation and dynamics in flow, which will be discussed later. 
The critical convergence length $L_e^c$ required for the development of RBC-FL is close to $25D$ (i.e., $L_e^c \approx 500$ $\mu m$ here) for both $C=1$ and $C=5$ suspensions, 
which is consistent with the previous investigation for $C=1$ \cite{Katanov_MVR_2015}. Nevertheless, $L_e^c$ becomes longer with increasing $C$, as $L_e^c$ is approximately $50D$ for $C = 20$. The decrease of $\delta$ from its initial thickness for 
the SC suspension makes it difficult to define $L_e^c$ due to a weak migration strength in this case. The development and dependence of RBC-FL thickness on $C$ is determined by the migration and interactions between RBCs as cellular core forms.

\begin{figure*}[!h]
	\centering
	\includegraphics*[width=1.0\textwidth]{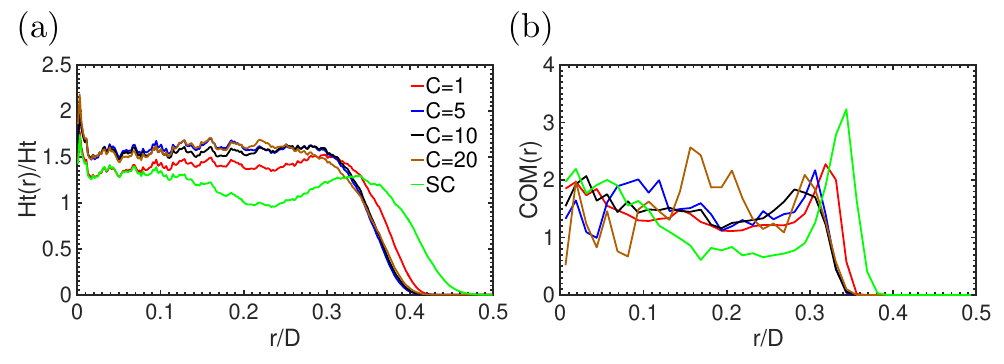}
	\caption{RBC distribution as a function of radial position within the tube. (a) Local hematocrit $H_{t}(r)/H_t$ profile for $C=1$, $5$, $10$, $20$ and SD suspensions. (b) Center-of-mass (COM) distribution of RBCs. 
		The data are collected after the flow has fully developed for $L_e>50D$. $H_{t}=30\%$, $D=20$ $\mu m$, and $Ca =  0.61$ ($\overline{\dot{\gamma}} = 377.4$ $s^{-1}$). }
	\label{fig:HtDis}
\end{figure*}

\subsection{Structure and mobility of the flowing RBC suspension}
\label{sec:struct} 

To better understand the dependence of final RBC-FL thickness $\delta_f$ on the viscosity contrast, it is instructive to take a look at the structural properties of RBC suspensions after the flow has fully developed. 
The distribution of RBCs within the tube can be characterized by local hematocrit $H_{t}(r)$ obtained from simulations through spatial averaging of the density of fluid particles inside RBC membranes. Figure \ref{fig:HtDis}(a) 
shows the normalized local hematocrit $H_{t}(r)/H_{t}$ within the tube for various viscosity contrasts. All $H_{t}(r)$ distributions contain a depletion zone near the wall, whose thickness is directly associated with $\delta_f$.
For instance, the difference in $\delta_f$ can clearly be seen for $C=1$, $C=5$ and SC suspensions. Inside the RBC-rich region, $H_{t}(r)$ is nearly uniform with
a small peak near the center, which is typical for small tube diameters \cite{Lei_BFT_2013}. For $C \geq 5$, the $H_{t}(r)$ distributions are nearly independent
of the viscosity contrast.

Radial cell density can also be characterized by the distribution of RBC centers of mass denoted as $COM(r)$ and shown in Fig.~\ref{fig:HtDis}(b). $COM$
distributions have a peak near the RBC-FL. As the tube center is approached, $COM(r)$ first decreases and then slightly increases for both $C=1$ and $C=5$
suspensions. As $C$ is increased, spatial inhomogeneity in $COM$ distribution becomes stronger. This is related to the RBC dynamics in flow and a decrease
in radial migration of RBCs with increasing $C$, which will be discussed later. The SC suspension shows the strongest variations in $COM(r)$, which is
consistent with the $H_{t}(r)$ distribution in Fig.~\ref{fig:HtDis}(a).

\begin{figure*}[!h]
	\centering
	\includegraphics*[width=1.0\textwidth]{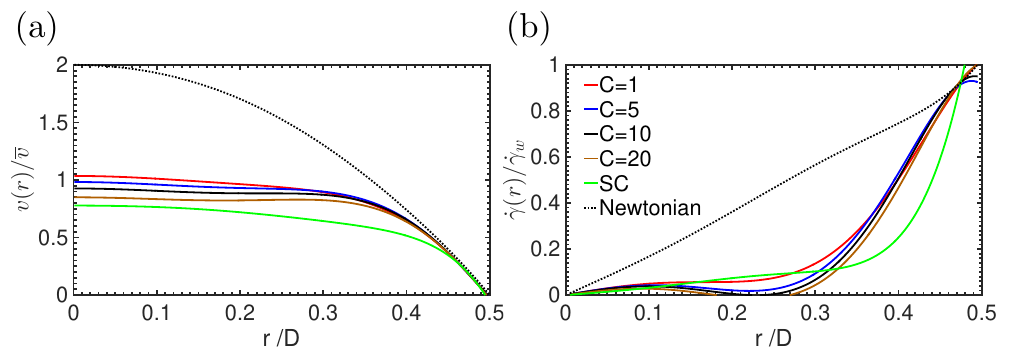}
	\caption{(a) Normalized velocity profile $v/\overline{v}$ and (b) local shear rate $\dot{\gamma}(r)/ \dot{\gamma}_w$ as a function of radial position within the tube for various RBC suspensions. 
		$H_{t}=30\%$, $D=20$ $\mu m$, and $Ca =  0.61$ ($\overline{\dot{\gamma}} = 377.4$ $s^{-1}$). }
	\label{fig:VelDis}
\end{figure*}

Figure \ref{fig:VelDis}(a) presents flow velocity profiles $v(r)$ normalized by the average velocity $\overline{v}$ for different RBC suspensions and blood
plasma (i.e., for a Newtonian fluid). All velocity profiles for RBC suspensions are flattened at the tube center due to the presence of the cellular core.
The velocity profiles do not exhibit large differences for various $C$ values and overlap with the Newtonian-fluid case only in the RBC-FL. An increase
of $C$ results in slight flattening of $v(r)$ near the tube center. Figure \ref{fig:VelDis}(b) shows radial profiles of local shear rates $\dot{\gamma}(r)$
normalized by the wall shear rate $\dot{\gamma}_w = 8\overline{\dot{\gamma}}$, which are computed from the velocity profiles $v(r)$ in Fig.~\ref{fig:VelDis}(a). Inside the cellular core,
$\dot{\gamma}(r)$ for all RBC suspensions is close to zero (i.e., plug flow) and much less than that for the Newtonian case. Within the RBC-FL, 
$\dot{\gamma}(r)$ quickly increases from nearly zero to the wall shear rate. The SC suspension exhibits a steeper increase in $\dot{\gamma}(r)$
within the RBC-FL in comparison to soft-RBC suspensions.

\begin{figure*}[!h]
	\centering
	\includegraphics*[width=1.0\textwidth]{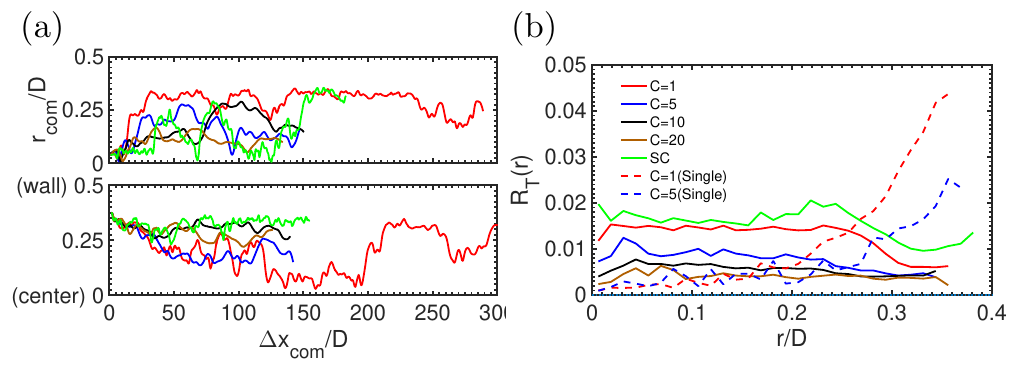}
	\caption{Characteristics of RBC mobility perpendicular to the flow direction. (a) Trajectories of selected individual cells whose initial positions are closer to the tube center (top) and closer to the wall (bottom). 
		(b) Distributions of the dimensionless lateral mobility coefficient $R_{T}(r)$ as a function of radial position within the tube. $H_{t}=30\%$, $D=20$ $\mu m$, and $Ca =  0.61$ 
		($\overline{\dot{\gamma}} = 377.4$ $s^{-1}$).}
	\label{fig:VrDis}
\end{figure*}

\begin{figure*}[!h]
	\centering
	\includegraphics*[width=0.6\textwidth]{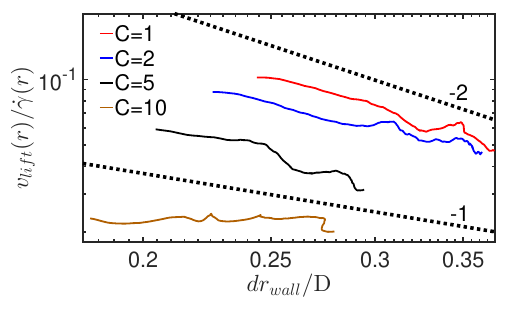}
	\caption{Lift velocity $v_l$ normalized by local shear rate $\dot{\gamma}(r)$ for a single RBC migrating away from the wall 
		as a function of the distance $d_w$ from the wall to the cell's COM in tube flow with $D=20$ $\mu m$. Dashed lines indicate the power-law functions of $d_w^{-1}$ and $d_w^{-2}$ 
		in the log-log plot.}
	\label{fig:MigrationSingle}
\end{figure*}

Even though fully developed velocity profiles are stable, RBCs within the cellular core are mobile, as can be seen from the trajectories of selected individual
cells in Fig.~\ref{fig:VrDis}(a). Thus, cells can migrate between different fluid layers laterally, even after the RBC-FL has fully developed ($L_e>50D$).
To characterize the lateral mobility of RBCs, we define a dimensionless lateral mobility coefficient $R_{T}=|\Delta r_{com}| / \Delta x_{com}$,
where the absolute value of the cell velocity in radial direction, $\Delta r_{com}/\Delta t$, is normalized by its translational velocity along
the x direction, $\Delta x_{com}/\Delta t$ (here $\Delta t \approx 8$ $ms$). $R_{T}(r)$ is shown in Fig.~\ref{fig:VrDis}(b) and can be interpreted as a measure of cell-cell collisions.
$R_T(r)$ remains nearly constant within the cellular core and decreases slightly near the RBC-FL. Interestingly, $R_{T}(r)$ or fluctuations in lateral
cell motion become smaller as $C$ increases. Nevertheless, a large enough increase in $C$ should eventually lead to an increase in $R_T(r)$, as for
the SC suspension, where $R_T(r)$ values are slightly larger than those for $C=1$. For comparison, we have also performed simulations of a single RBC
migrating from the wall to the tube center for $C=1$ and $C=5$ (the dashed lines in Fig.~\ref{fig:VrDis}(b)). For a single cell,  lateral migration 
due to the hydrodynamic lift force near the wall is much faster than that in a suspension, because cell-cell collisions are not present. However, in the tube center,
cell-cell interactions within RBC suspensions enhance lateral migration of cells in comparison to the case of a single RBC. 

Note that $R_T(r)$ of a single RBC for $C=1$ in Fig.~\ref{fig:VrDis}(b) is larger than for $C=5$, indicating a stronger lateral migration. 
Figure \ref{fig:MigrationSingle} presents the lift velocity $v_l$ of a single migrating RBC normalized by local shear rate $\dot{\gamma}(r)$ for $C \in [1,10]$. 
Clearly, the RBC with $C=1$ migrates faster than that with $C=5$. This result is consistent with previous simulation studies \cite{Messlinger_DRH_2009,Narsimhan_CGT_2013}, 
where the lift velocity on single RBCs in pure shear flow has been found to decrease with increasing $C$. The ratio $v_l/\dot{\gamma}(r)$ is expected to 
be proportional to $1/d_w^\alpha$, where $d_w$ is the distance from the wall to the cell's COM. We find that $\alpha \simeq 2$ for $C= \leq 2$, and $1 < \alpha <2$ for $C =5$, in agreement 
with experimental measurements \cite{Grandchamp_LSD_2013,Abkarian_DVF_2005,Abkarian_TTU_2002}. For the RBC with $C = 10$, $\alpha$ is nearly zero. 
Note that these simulations are performed in a tube with diameter $D=20$ $\mu m$ (${D_r/D}=0.33$), representing a rather strong confinement with varying 
local shear rates. Furthermore, initial cell migration might be affected by the flow development, as we start from no-flow conditions. 

The faster migration velocity of RBCs 
for $C=1$ in comparison with $C=5$ is due to differences in cell dynamics for different $C$. At high enough shear rates (e.g., near the wall), RBCs generally 
exhibit a tank-treading-like motion with a preferred alignment in flow for $C=1$, while a tumbling-like dynamics with multi-lobed shapes is found for $C=5$ 
\cite{Lanotte_RCM_2016,Mauer_FIT_2018}. The tank-treading dynamics of RBC membrane near a wall leads to a larger lift force than for the tumbling 
dynamics \cite{Pozrikidis_OMS_2005,Olla_LTT_1997}. Note that the slower migration velocity of RBCs for $C=5$ in comparison to $C=1$ is consistent 
with a slower development of the RBC-FL for $C=5$ in Fig.~\ref{fig:CFL}(a).

\begin{figure*}[!h]
	\centering
	\includegraphics*[width=0.8\textwidth]{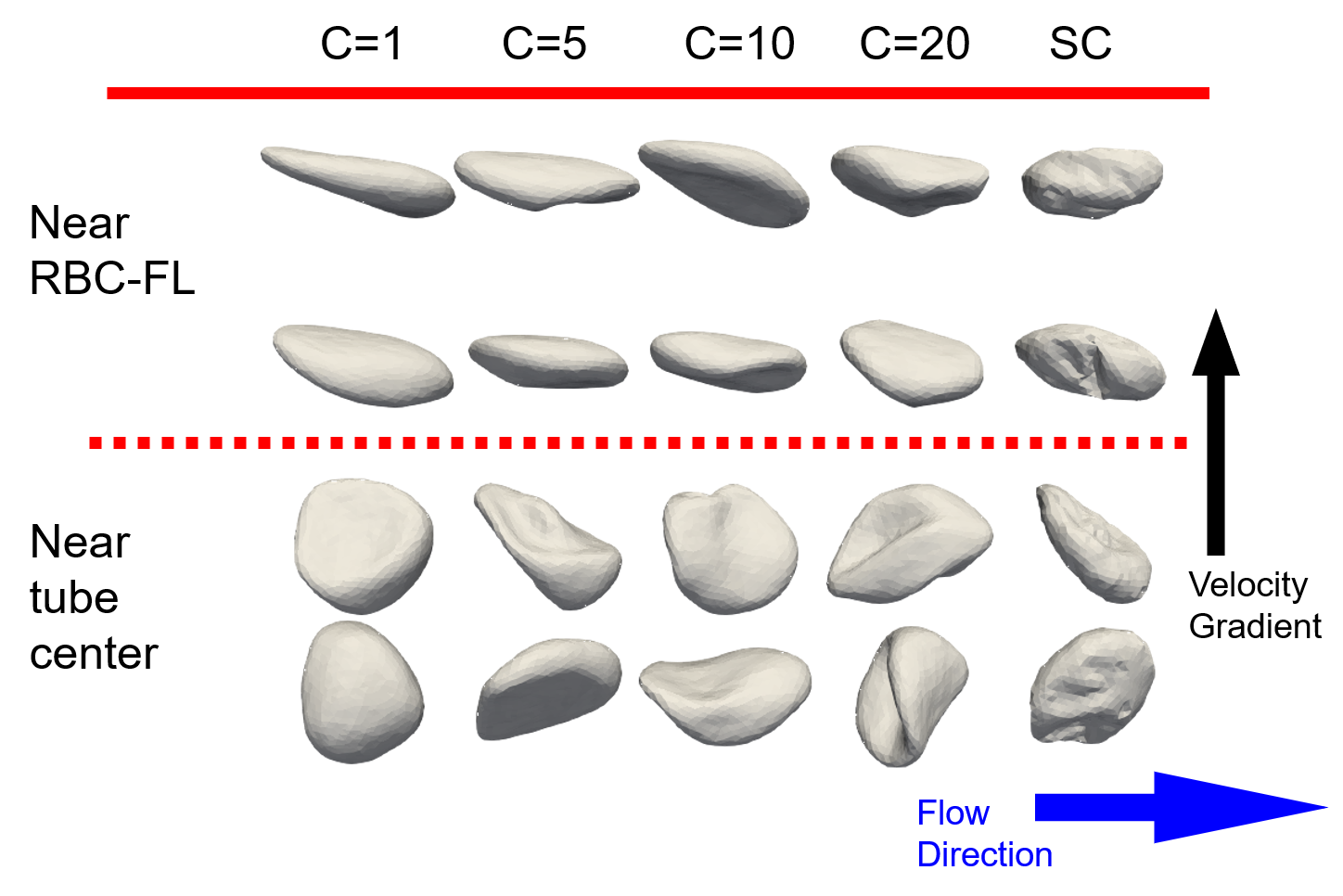}
	\caption{Representative snapshots of RBC shapes near the edge of the cellular core and in the tube center for different RBC suspensions.
          Each column denotes a specific $C$ value. $H_{t}=30\%$, $D=20$ $\mu m$, and $Ca =  0.61$ ($\overline{\dot{\gamma}} = 377.4$ $s^{-1}$).}
	\label{fig:shapeGallery}
\end{figure*}

\begin{figure*}[!h]
	\centering
	\includegraphics*[width=1.0\textwidth]{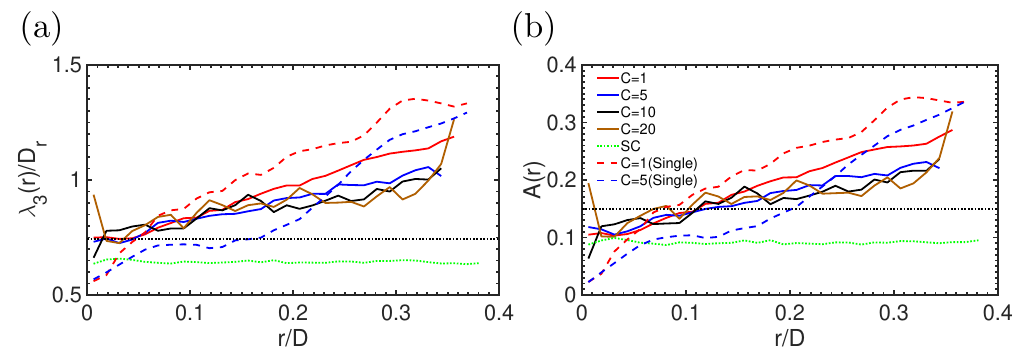}
	\caption{Characterization of RBC shapes in blood flow. (a) The largest eigenvalue $\lambda_{3}$ of the gyration tensor, showing relative
          RBC stretching in the cellular core. (b) Cell asphericity $A$. Different colors mark different RBC suspensions. The horizontal dashed
          line corresponds to the RBC shape at rest. The red and blue dashed lines represent the shapes of a single cell in tube flow.
          $H_{t}=30\%$, $D=20$ $\mu m$, and $Ca =  0.61$ ($\overline{\dot{\gamma}} = 377.4$ $s^{-1}$).}
	\label{fig:shape}
\end{figure*}

\subsection{Shape and dynamics of single cells inside the RBC core}
\label{sec:shape}

RBC dynamics inside the cellular core involves frequent changes in the shape, orientation, and membrane tank-treading motion, which often cannot
be decoupled completely. Figure \ref{fig:shapeGallery} shows several representative snapshots of RBCs near the RBC-FL and in the tube center.
To better understand the behavior of single RBCs within the cellular core, we look at several cell characteristics. Shape changes of RBCs are
quantified by the asphericity $A= \left[{\left({\lambda }_1-{\lambda }_2\right)}^2+{\left({\lambda }_2-{\lambda }_3\right)}^2+{\left({\lambda }_1
 -{\lambda }_3\right)}^2\right] / \left(2{R_g}^4 \right)$, where ${R_g}^2={\lambda }_1+{\lambda }_2+{\lambda }_3$ is the gyration radius squared and
${\lambda }_1\mathrm{\le }{\lambda }_2\mathrm{\le }{\lambda }_3$ are the eigenvalues of the gyration tensor. For a biconcave RBC shape at rest,
$A=0.15$, while $A=0$ corresponds to a sphere and $A=1$ to a long thin rod. For example, the largest eigenvalue ${\lambda}_3$ can
be interpreted as the extension of a RBC in the flow direction (${\lambda }_3=4.77\mu m$ for a RBC at rest).

Figure \ref{fig:shape}(a) shows that RBCs are significantly stretched in most locations within the cellular core region, except near the tube center.
Close to the center, different shapes with $A\mathrm{<0.15}$ are generally observed (Fig.~\ref{fig:shape}(b)), whose representative
snapshots are shown in Fig.~\ref{fig:shapeGallery}. Both ${\lambda }_3$ and $A$ in Fig.~\ref{fig:shape} increase nearly linearly
from the tube center to the RBC-FL. At the edge of the cellular core, elongated slipper shapes prevail, as shown in Fig.~\ref{fig:shapeGallery}.
SC cells show a crumpled configuration, which is due to not only fluid-flow stresses, but also residual elastic stresses within the membrane, 
since the RBC stress-free shape corresponds to an oblate spheroid with a reduced volume of 0.96 and the SC cells have a high shear modulus. 
Therefore, stiffened cells even without flow show some degree of membrane roughness in comparison with the smooth surface of soft RBCs. 
Nevertheless, the overall rest shape of SC cells remains biconcave. The shape characteristics of SC cells in Fig.~\ref{fig:shape} are nearly 
independent of the radial position within the tube, indicating that cell deformation can nearly be neglected. As $C$ increases from $1$ to $5$, 
${\lambda }_3$ decreases and its slope reduces as well, indicating that RBCs at $C=5$ are stretched less than those at $C=1$. Suspensions 
with $C>5$ show similar shapes as those for $C=5$. In comparison to RBCs in a suspension, single cells in tube flow have a larger (smaller) 
elongation than for $C=1$ ($C=5$). 

\begin{figure*}[!h]
	\centering
	\includegraphics*[width=0.7\textwidth]{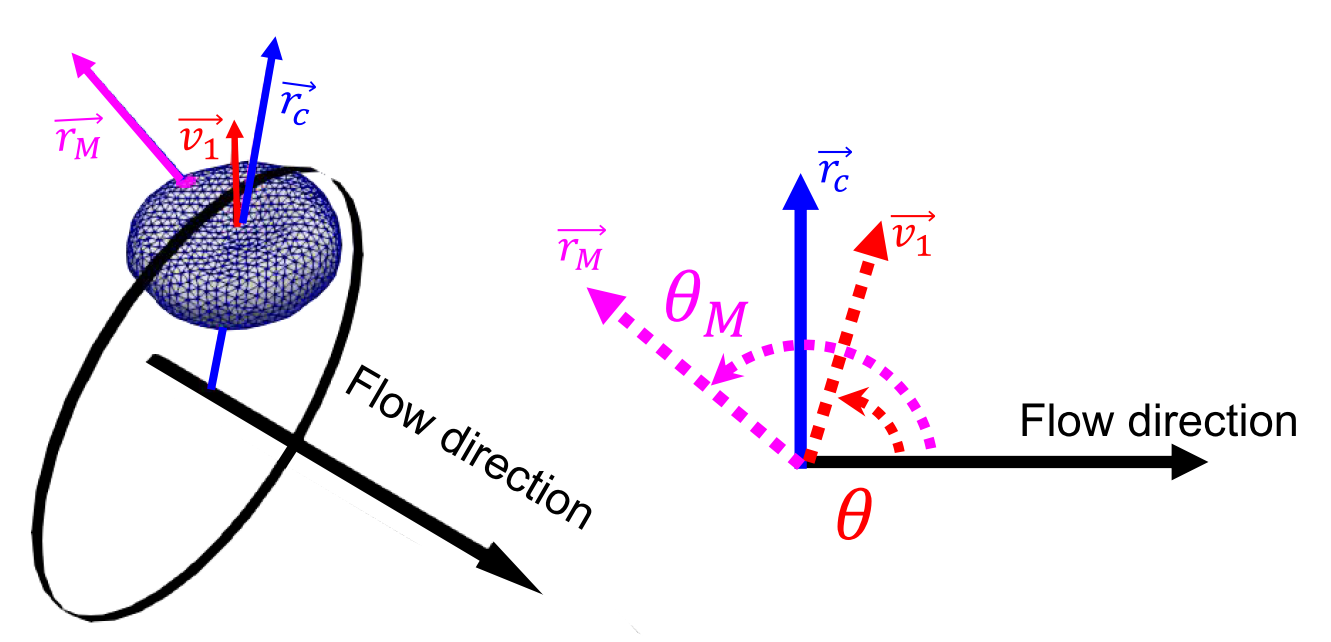}
	\caption{Definition of the inclination angle $\theta$ and the material angle $\theta_{M}$. Eigenvector $\vec{v_{1}}$ of the gyration tensor
          (red arrow), the vector $\vec{r_{c}}$ from the tube central axis to the cell's center of mass (blue arrow), and the material vector
          $\vec{r_{M}}$ from the cell's center of mass to a specific material point (pink arrow) are shown.}
	\label{fig:DefineAngles}
\end{figure*}

\begin{figure*}[!h]
	\centering
	\includegraphics*[width=1.0\textwidth]{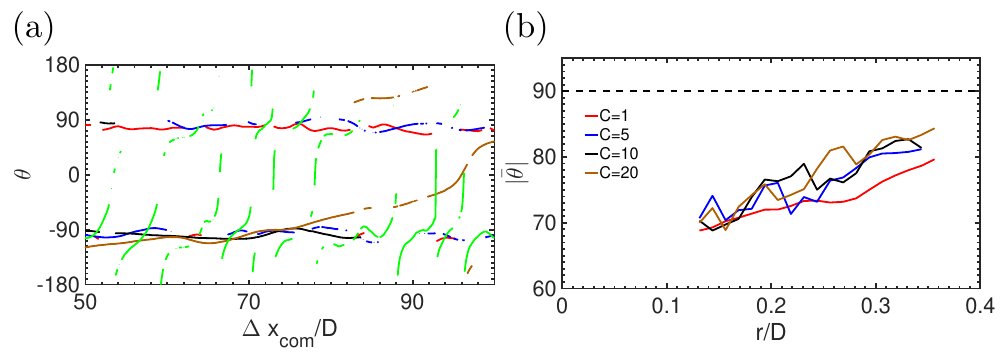}
	\caption{(a) Orientation angle $\theta$ of a selected cell as a function of RBC displacement. (b) Dependence of the average inclination angle $|\overline{\theta}|$
          on the radial position within the tube flow. Different suspensions are represented by different colors: $C=1$ (red), $C=5$ (blue), $C=10$ (black),
          $C=20$ (pink), and SC suspension (green). $H_{t}=30\%$, $D=20$ $\mu m$, and $Ca =  0.61$ ($\overline{\dot{\gamma}} = 377.4$ $s^{-1}$).}
	\label{fig:Theta}
\end{figure*}

In addition to deformation in flow, cell orientation changes and membrane exhibits a tank-treading motion. To quantify RBC orientation with a varying
shape, we include in the analysis only the shapes with an asphericity larger than that of a biconcave shape at rest ($A>0.15$), where a well-defined
axis from the eigenvector $\vec{v_{1}}$ corresponding to $\lambda_{1}$ can always be obtained. $\vec{v_{1}}$ often does not lie within
the flow-velocity gradient plane, due to cell-cell interactions and complex cell relaxation under fluid stresses. To simplify the quantification of
cell dynamics, we define an inclination angle $\theta$ (marked in red in Fig.~\ref{fig:DefineAngles}) as the angle between the vector $\vec{v_{1}}$
and the flow direction $x$ within the flow-velocity gradient plane (i.e. within $x-\vec{r_{c}}$ plane, where $\vec{r_{c}}$ is the vector from the tube
central axis to the cell's center of mass). Figure \ref{fig:Theta}(a) shows $\theta$ as a function of cell position in the flow direction for one
selected cell from each suspension. $\theta$ may frequently exhibit discontinuous jumps of 180 degrees due to the symmetric disk-like rest shape,
so that even small deformations can cause a reversal of the $\vec{v_{1}}$ vector. Despite these jumps, RBC tumbling in the SC suspension can clearly
be identified from the green trajectories in Fig.~\ref{fig:Theta}(a). Tumbling motion is infrequent for soft-RBC suspensions, where jumps between
$+90^o$ and $-90^o$ are found instead and the departure of $\theta$ from these two values is generally within $30$ degrees. Rare tumbling of RBCs for
$C=20$ suspension can be seen by the brown line in Fig.~\ref{fig:Theta}(a), but the tumbling period is extremely long ($>120D$). Thus, cell-cell interactions
in a crowded environment strongly hinder solid-like tumbling motion and RBCs are forced to relax fluid stresses via shape deformation instead.

The average orientation angle of RBCs as a function of radial position can be tracked through the absolute value of $|\overline{\theta}|$, as shown in Fig.~\ref {fig:Theta}(b).
Near the RBC-FL, the average angle $\overline{\theta}$ approaches $90^o$, so that RBCs are aligned along the flow direction. $\overline{\theta}$ decreases toward the tube center, and near
the center the orientation angle is often not well defined as RBCs attain shapes with $A<0.15$. The average angle $\overline{\theta}$ for $C \geq 5$ is slightly larger than
for $C=1$, indicating that RBCs are more aligned with the flow for large $C$ values, which can also be seen in Fig.~\ref{fig:shapeGallery}.

\begin{figure*}[!h]
	\centering
	\includegraphics*[width=1.0\textwidth]{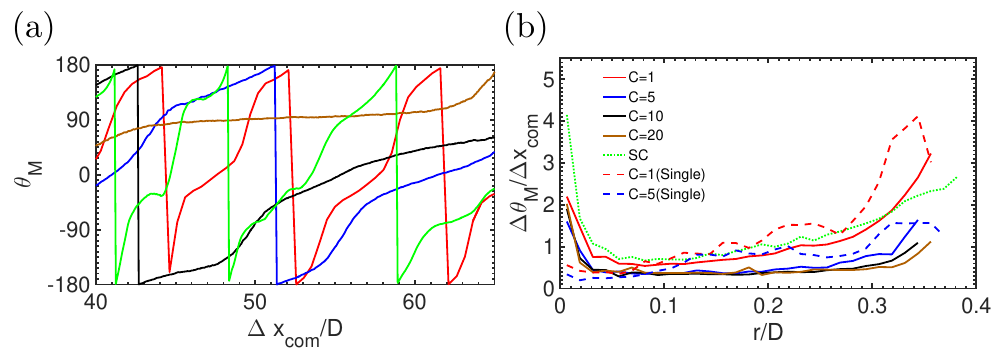}
	\caption{RBC membrane motion characterized by $\theta_{M}$. (a) $\theta_{M}$ evolution for a single selected RBC in flow. (b) Average change
          in $\theta_{M}$ obtained as the ratio  $\Delta \theta_{M}/ \Delta x_{com}$ as a function of $r$. $H_{t}=30\%$, $D=20$ $\mu m$, and
          $Ca =  0.61$ ($\overline{\dot{\gamma}} = 377.4$ $s^{-1}$).}
	\label{fig:ThetaM}
\end{figure*}

Strong dynamic changes in RBC shape significantly complicate cell orientation analysis and do not always allow decoupling between tumbling motion
with shape rotation and tank-treading motion with membrane circulation. To analyze relative membrane motion, we monitor the material angle
$\theta_{M}$ defined as the angle between the material vector $\vec{r_{M}}$ and the flow direction, see Fig.~\ref{fig:DefineAngles}. Note that we cannot
fully distinguish between tank-treading and tumbling motion by using $\theta_{M}$. Changes in $\theta_{M}$ for a selected RBC are shown in
Fig.~\ref{fig:ThetaM}(a) for various viscosity contrasts. $\theta_{M}$ exhibits a nearly monotonic increase with increasing $C$. For $C=1$, a smooth dependence of
$\theta_{M}$ is observed, indicating continuous tank-treading motion of the membrane. For $C \geq 5$, an increase in $\theta_{M}$ is slower than for $C=1$,
and membrane rotation shows short frequent pauses followed by periods of rapid $\theta_{M}$ increase.

Membrane motion can be quantified further by an average change in $\theta_{M}$ as RBCs flow, which is defined as the ratio $\Delta \theta_{M}/ \Delta x_{com}$
between the change in $\theta_{M}$ and the corresponding change in $x_{com}$ calculated for a fixed time interval $\Delta t \approx 8$ $ms$. Figure \ref{fig:ThetaM}(b) presents $\Delta \theta_{M}/ \Delta x_{com}$, where
the rotation tendency is most prominent near the RBC-FL and decreases toward the tube center. A slight enhancement of $\Delta \theta_{M}/\Delta x_{com}$ can
be observed near the tube center. Large $C$ values suppress membrane rotation, since a large difference in $\Delta \theta_{M}/ \Delta x_{com}$ is observed between $C=1$ and $C=5$ suspensions.
The dependence of $\Delta \theta_{M}/ \Delta x_{com}$ is similar for all $C \geq 5$ suspensions. In comparison to a single cell dynamics in tube flow,
the membrane rotation for cell suspensions is suppressed by the surrounding cells. 

\begin{table}[]
	\centering
	\caption{Relative viscosity $\eta_{rel}$ and final RBC-FL thickness $\delta_f$ for various flow conditions characterized by $Ca$. $D=20$ $\mu m$ and $H_{t}=30\%$.}
	\begin{tabular}{c|c|c|c|c}
		$Ca$                        & 0.61   & 0.31   & 0.14   & 0.06   \\ \hline \hline
		$\eta_{rel}^{C=1}$            & 1.27  & 1.33  & 1.46  & 1.79  \\ \hline
		$\eta_{rel}^{C=5}$            & 1.30  & 1.41  & 1.53  & 1.86  \\ \hline
	        $\eta_{rel}^{C=5}$/ $\eta_{rel}^{C=1}$-1   & 2.15\% & 6.25\% & 4.61\% & 3.50\%  \\ \hline
                $\delta_f^{C=1}$ $[\mu m]$ &  2.77 & 2.79 & 2.66 & 2.42 \\ \hline
                 $\delta_f^{C=5}$ $[\mu m]$ &  3.12 & 2.90 & 2.72 & 2.50 \\ \hline
	\end{tabular}
\label{tab:table_f}
\end{table}

\subsection{Dependence of the RBC-FL thickness on other flow parameters}
\label{sec:other} 

To study the dependence of RBC-FL thickness on other flow parameters, we compare suspensions with $C=1$ and $C=5$ for different flow conditions
$Ca\in[0.06,0.61]$ ($\overline{\dot{\gamma}}\in[37,378]\ s^{-1}$), hematocrits $H_{t}\in[15,45]\ \%$, and tube diameters $D\in[10,40]\ \mu m$. 
Table \ref{tab:table_f} presents relative viscosity $\eta_{rel}$ and final RBC-FL thickness $\delta_f$ for various $Ca$, where $\eta_{rel}$ increases
as $Ca$ decreases. Furthermore, $\eta_{rel}$ for $C=5$ suspension is only slightly larger than $\eta_{rel}$ for $C=1$ suspension and the difference
is most prominent at intermediate flow rates. The converged RBC-FL thickness $\delta_f$ decreases as the flow slows down. The $C=1$ suspension
has a smaller RBC-FL thickness than the $C=5$ suspension with a difference of about $350$ $nm$ at $Ca=0.61$ and about $100$ $nm$ at $Ca=0.06$.

\begin{table}[]
\centering
\caption{Relative viscosity $\eta_{rel}$ and final RBC-FL thickness $\delta_f$ for various $H_{t}$ values. $D=20$ $\mu m$ and
  $Ca =  0.61$ ($\overline{\dot{\gamma}} = 377.4$ $s^{-1}$).}
	\begin{tabular}{c|c|c|c}
		$H_{t}$                    & 15\%     & 30\%     & 45\%     \\ \hline \hline
		$\eta_{rel}^{C=1}$          & 1.13  & 1.27  & 1.72  \\ \hline
		$\eta_{rel}^{C=5}$          & 1.16  & 1.30  & 1.87  \\ \hline
		$\eta_{rel}^{C=5}$/ $\eta_{rel}^{C=1}$-1   & 2.59\% & 2.15\% & 9.09\% \\ \hline
                $\delta_f^{C=1}$ $[\mu m]$ &  1.45 & 2.77 & 3.65  \\ \hline
                 $\delta_f^{C=5}$ $[\mu m]$ &  1.65 & 3.12 & 3.88 \\ \hline
	\end{tabular}
\label{tab:table_Ht}%
\end{table}

\begin{figure*}[!h]
	\centering
	\includegraphics*[width=0.6\textwidth]{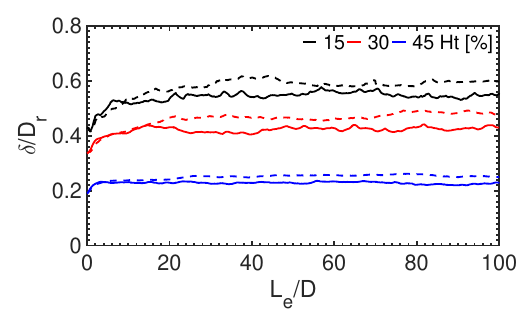}
	\caption{Development of the RBC-FL for $C=1$ (solid lines) and $C=5$ (dashed lines) suspensions and different hematocrits
          $H_{t}=15\%$ (black), $H_{t}=30\%$ (red), and $H_{t}=45\%$ (blue). $D=20$ $\mu m$ and
        $Ca =  0.61$ ($\overline{\dot{\gamma}} = 377.4$ $s^{-1}$).}
	\label{fig:cflScanHt}
\end{figure*}

Table \ref{tab:table_Ht} shows that $\eta_{rel}$ increases as $H_{t}$ increases. The difference in $\eta_{rel}$ caused by $C$ is most prominent
for dense suspensions. The development of RBC-FL for different $H_t$ values is shown in Fig.~\ref{fig:cflScanHt}, where $\delta_f$ decreases
for increasing $H_t$, consistently with the increase of $\eta_{rel}$. $\delta_f$ is nearly the same for $C=1$ and $C=5$ suspensions. The difference
between $C=1$ and $C=5$ for $\eta_{rel}$ and $\delta_f$ remains similar for hematocrits $H_{t}=15\%$, $30\%$, and $45\%$. 

\begin{figure*}[!h]
	\centering
	\includegraphics*[width=0.6\textwidth]{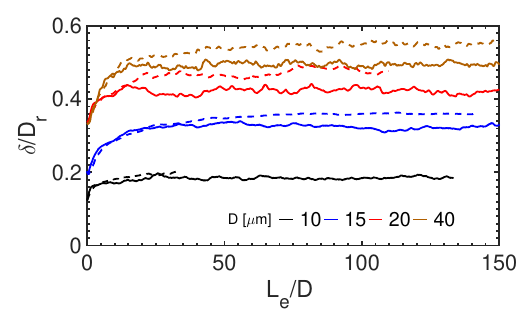}
	\caption{Development of the RBC-FL for $C=1$ (solid lines) and $C=5$ (dashed lines) suspensions and tube diameters $D=10$ $\mu m$ (black), 
	    $D=15$ $\mu m$ (blue),	$D=20$ $\mu m$ (red), and $D=40$ $\mu m$ (brown). $H_{t}=30\%$ and $Ca =  0.61$ ($\overline{\dot{\gamma}} = 377.4$ $s^{-1}$). }
	\label{fig:cflScanD}
\end{figure*}

Simulations of flowing RBC suspensions at $H_{t}=30\%$ for different tube diameters $D$ with a constant $Ca =  0.61$ show that $\eta_{rel}$ decreases
as $D$ decreases (see Table \ref{tab:table_D}), which is consistent with experiments \cite{Fahraeus_VOB_1931}. The difference in $\eta_{rel}$ for
the two $C$ values is most prominent for the large diameter $D=40$ $\mu m$.
Consistently, $\delta_f$ increases as $D$ increases, as shown in Fig.~\ref{fig:cflScanD}. This trend
may depend on $Ca$, which affects dynamics of RBCs near the RBC-FL. The difference between $C=1$ and $C=5$ suspensions is similar for
$D=40$ $\mu m$ and $D=20$ $\mu m$, but nearly disappears for $D=10$ $\mu m$.

\begin{table}[]
	\centering
	\caption{Relative viscosity $\eta_{rel}$ and final RBC-FL thickness $\delta_f$ for various tube diameters $D$. $H_{t}=30\%$ and
        $Ca =  0.61$ ($\overline{\dot{\gamma}} = 377.4$ $s^{-1}$).}
	\begin{tabular}{c|c|c|c|c}
		$D$ {[}$\mu m${]}            & 40     & 20     & 15     & 10     \\ \hline \hline
		$\eta_{rel}^{C=1}$            & 1.39  & 1.27  & 1.25  & 1.12  \\ \hline
		$\eta_{rel}^{C=5}$            & 1.51  & 1.30  & 1.21  & 1.16  \\ \hline
		$\eta_{rel}^{C=5}$/ $\eta_{rel}^{C=1}$-1    & 8.5\%  & 2.2\%  & -3.2\% & 3.2\%  \\ \hline
                $\delta_f^{C=1}$ $[\mu m]$ &  3.19 & 2.77 & 2.18 & 1.21 \\ \hline
                 $\delta_f^{C=5}$ $[\mu m]$ &  3.53 & 3.12 & 2.34 & 1.20 \\ \hline
	\end{tabular}
\label{tab:table_D}
\end{table}

\section{Summary and discussion}
\label{sec:summary}

The main focus of our study is the effect of the viscosity ratio $C$ on the behavior of RBC suspensions. Interestingly, $C\in[1,20]$ has only a very weak effect
on the resistance of converged flow, quantified by the relative viscosity $\eta_{rel}$. We have systematically analyzed the RBC-FL thickness, which is closely 
related to the flow resistance. When $C$ increases from $1$ to $5$, the change in $\eta_{rel}$ is within $2\%$, while $\delta_f$
increases by about $10-15\%$ (or by approximately $300$ $nm$). A further increase of $C$ from $5$ to $20$ leads to an increase of $\eta_{rel}$ by about $6\%$, while $\delta_f$ stays
nearly unaffected. Thus, a larger viscosity of the RBC core for $C\geq 5$ in comparison to $C=1$ is complemented by a larger RBC-FL thickness, such
that the overall flow resistance remains nearly independent of $C$.

The development of RBC-FL is governed by the two main mechanisms \cite{Katanov_MVR_2015}: (i) RBC migration away from the wall due to the hydrodynamic lift force and (ii) the
dispersion of RBCs within the cellular core due to cell-cell interactions in flow. The migration of RBCs away from the wall can be attributed
to their shape deformation and membrane dynamics. For instance, RBCs with $C=1$ migrate faster than those with $C=5$, which is consistent with a slower development 
of the RBC-FL for $C=5$ in Fig.~\ref{fig:CFL}(a). This results from the fact that both shape deformation and membrane
dynamics are suppressed as $C$ increases from $1$ to $5$, since a larger internal viscosity dampens RBC shape changes and dynamics. The importance of
cell deformability for migration and the formation of RBC-FL is further illustrated by the results for SC suspension. Stiffened RBCs exhibit
tumbling dynamics, which leads to a weak lift force, small RBC-FL thickness, and a large flow resistance.

The dispersion of RBCs within the cellular core counterbalances the lift force and is governed by local shear rate, shape deformations, and membrane
dynamics. Both shape deformations and membrane dynamics are attenuated for $C=5$ in comparison to $C=1$. For example, the shape asphericity $A$
for $C=5$ is smaller than for $C=1$, indicating that RBCs are less stretched. Furthermore, membrane dynamics for $C=5$ is slower than for
$C=1$, because membrane tank-treading is more pronounced for the low viscosity contrast. This leads to a weaker hydrodynamic repulsion
between RBCs in the $C=5$ suspension, whose origin is similar to the lift force near a wall. To further confirm that
the dispersion of RBCs is larger for $C=1$ than for $C=5$, we have performed several simulations of the collision of two cells in tube flow.
After subtracting the migration effect, we find that the collision of two RBCs leads to a larger change in lateral cell displacement for
$C=1$ than for $C=5$, indicating a stronger dispersion effect for $C=1$. Even though cell collisions in dense suspensions involve more complex
multicellular interactions, the lateral mobility coefficient $R_{T}$ in Fig.~\ref{fig:VrDis}(b) is smaller for $C=5$ than for $C=1$, which
is consistent with the simulations of binary collisions. Therefore, the cellular
core for $C=5$ remains more compactly packed than for $C=1$, which is documented by a slightly thicker RBC-FL. The importance of RBC
dynamics in flow is further emphasized by the behavior of stiffened RBCs. Hardened cells exhibit tumbling dynamics even in the cellular core,
which results in their significant dispersion within the tube and a thin RBC-FL.

The dependence of the converged RBC-FL thickness $\delta_f$ on the flow rate is consistent with the discussion of cell deformation and dynamics above,
such that a decrease in driving pressure gradient weakens cellular dynamics and the exerted lift force, resulting in a reduction of $\delta_f$ and an increase of flow resistance.
Thus, the flow resistance is larger in the venular part of microvasculature, where blood-flow velocities are significantly smaller than
those in the arteriolar part of the microvasculature. As expected, an increase in hematocrit leads to a decrease of $\delta_f$ due to increased
cell crowding, accompanied also by a slightly attenuated RBC dynamics. Note that the difference in $\eta_{rel}$ between $C=1$ and $C=5$ is more
pronounced for $H_{t}=45\%$ than for smaller $H_t$ values. However, this difference can likely be neglected, because microvascular hematocrit
values are generally smaller than $35\%$. A decrease of tube diameter $D$ causes more pronounced cell-cell and cell-wall interactions,
which result in a reduction of flow resistance, consistently with the Fahraeus-Lindqvist effect \cite{Fahraeus_VOB_1931,Pries_BVT_1992}.
Note that the viscosity ratio $C$ has a more significant effect on flow resistance for large tube diameters ($D>40$ $\mu m$), where $\eta_{rel}$
for $C=5$ can be larger by more than $10\%$ than for $C=1$. 

An interesting observation from our investigation is that the final RBC-FL thickness first increases with increasing
$C$ and then decreases for stiffened RBCs (this can be approximated by $C \to \infty$). As discussed above, the initial increase in $\delta_f$
for increasing $C$ is the main reason that the flow resistance is nearly unaffected by $C\in[1,20]$. In fact, this property of flowing blood allows
the maximization of hemoglobin content in RBCs, and therefore oxygen delivery, without negative effects on the flow resistance. Furthermore, 
the effect of $C$ in this range on blood-flow convergence is rather weak, such that RBC-FL convergence is obtained after a distance of about 
$25D-50D$. The flow-convergences distance can easily be compared with an average length of vessels ($0.5-1$ $mm$) within a branching 
network-like microvasculature \cite{Popel_MH_2005,Secomb_BFM_2017}, e.g. $L_e=50D=1$ $mm$ for $D=20$ $\mu m$. This means that a converged flow within 
the microvasculature can only be expected in small vessels such as capillaries, while in microvessels with a diameter larger than about $10-20$ $\mu m$,
blood flow would likely always correspond to a transient (non-converged) flow. Even though the dependence of $\eta_{rel}$ on $C\in[1,20]$ is nearly negligible, the structure and dynamics of RBC suspension are 
different for various $C$ values. These differences in flow behavior for different viscosity contrasts are likely to be important for the margination 
of particles (e.g., platelets, drug-delivery carriers) in blood flow as well as for partitioning of RBCs within a complex microvascular network.  
In particular, the slow flow convergence behind branching points implies that many flow properties, such as resistance, particle margination, 
and oxygen delivery, can be highly inhomogeneous in complex vessel networks, depending on vessel diameters, branching lengths, and distance from 
the previous branching point. These aspects of blood flow need still to be addressed in future research.   

\section*{Acknowledgements}
The authors gratefully acknowledge the computing time granted through JARA-HPC on the supercomputer JURECA \cite{jureca} at Forschungszentrum J\"ulich.


%

\end{document}